# Flowing cryogenic liquid target for terahertz wave generation

Yiwen E[1], Yuqi Cao[1,2], Fang Ling[1,3], and X.-C. Zhang[1, *]

[1]The Institute of Optics, University of Rochester, Rochester, NY, 14627 USA

[2]College of Control Science and Engineering, Zhejiang University, Hangzhou 310027, China

[3]College of Electronics and Information Engineering, Sichuan University, Chengdu, 610065 China

* Corresponding author: xi-cheng.zhang@rochester.edu

Terahertz wave emission from condensed matter excited by intense laser pulses not only reflects the details in laser-matter interaction but also offers bright terahertz wave sources. Flowing liquid targets possess the advantage of providing a fresh area for each laser pulse. To demonstrate a debris-free target under laser excitation, we investigate the use of liquid nitrogen as a target. By creating a flowing liquid nitrogen line in the ambient environment, we successfully observe broadband terahertz wave emission under short pulse excitation. Our cryogenic line is able to sustain the excitation of a high-repetition-rate (1 kHz) laser. The terahertz peak field emitted from liquid nitrogen is comparable to that from liquid water, yet a broader bandwidth is observed. This demonstration prompts new opportunities in choosing potential materials for studying terahertz wave generation process and in understanding laser-induced ionization in different liquids.

The rapid development of advanced laser technology provides great opportunities to study nonlinear processes in laser-matter interaction. In terahertz (THz) regime, there is an increasing demand for intense THz sources to enable fundamental studies in ultrafast phenomena,[1-3] such as hidden phase transition,[4] high harmonic generation,[5] alignment and orientation of molecules.[6] Currently, a THz field over MV/cm is attainable by employing nonlinear crystals.[7,8] Further improvement is limited by the optical damage under intense laser irradiation. Additionally, THz pulse energy above 50 mJ has been reported, with a solid metal target under-single-shot excitation by an intense laser pulse (60 J).[9] However, because of the contamination issue caused by debris as well as the target damage, solid targets are hardly applied to lasers with a high repetition rate (1 kHz). Therefore, developing a durable target is imperative to provide a solution for intense lasers. Liquid targets with a similar density as solid are the potential candidates due to their capability of providing a fresh area for each pulse. Moreover, liquid targets have been studied for decades in generating extreme ultraviolet and X-ray radiation.[10,11]

Since the first observation of THz wave generated from liquid water,[12,13] more experiments and discussions based on liquid targets for THz wave generation have been reported. Remarkably, THz field strength up to 0.2 MV/cm has been demonstrated by using a 200 μm water line as the target.[14] The broadband THz wave generated from liquid metal is also observed.[15] Also, the enhancement of THz wave emission induced by a pre-existing plasma in a double-pump excitation geometry is studied to reveal more details in laser-liquid interaction.[16,17] In contrast to an air plasma, the preference for subpicosecond excitation pulses suggests a different ionization processes in liquids, in which collisional ionization plays an important role in increasing the electron density.[18] On the other hand, unlike a solid target, the fluidity of liquids can support a continued operation without interruption. Nevertheless, there is still debris when laser intensity is high enough to create a "mist" from the target, which not only contaminates the optics nearby, but also absorbs and scatters the signal. It should be noted that while gas is a good debris-free target, it hardly supports a high-density plasma because of the relatively low molecular density.

To demonstrate a debris-free target for THz wave emission, we create a free-standing, flowing liquid nitrogen ($LN_2$) line in the ambient environment. Applying single color excitation, coherent THz emission is detected in the forward direction by electro-optic sampling (EOS). Compared with the signal from liquid water under the same excitation condition, the signal from $LN_2$ has a comparable peak field but a broader bandwidth, which is attributed to the low absorption of $LN_2$ at higher THz frequency. The $LN_2$ target can support the excitation by laser pulses with a 1 kHz repetition rate without interruption.

**Fig. 1a** shows the apparatus for creating a gravity-driving, free-standing liquid line of $LN_2$. It consists of a syringe, a flask and an insulating layer. Specifically, a metal syringe is used for guiding the flow of liquid nitrogen and it is immersed in a flask filled with $LN_2$ for maintaining the cryogenic temperature. Outside the flask, a thick insulating layer is employed to resist the heat transition between the flask and the ambient environment. The volume of the flask is 600 ml. While filling the flask with $LN_2$, the syringe is blocked on purpose at the beginning until the setup is cooled down. Benefiting from the high thermal conductivity of the metal syringe, the setup reaches a thermal balance in a few minutes. After removing the block, a small amount of vaporized nitrogen gas is first ejected out from the syringe needle. Then, a steady liquid line is formed.

**Fig. 1b** is a photo taken by a CCD camera showing the flowing $LN_2$ line. The inner diameter of the syringe needle is about 410 μm. From the photo, the diameter of the flow is estimated to be 400 ± 5 μm. The liquid line shows high transparency indicating a smooth surface and a stable flow. Also, the video in **Supplementary S1** clearly shows the flowing $LN_2$ line. For comparison, **Fig. 1c** shows the ejection of a gas-liquid mixture when the flask is connected to a Dewar directly. The pressure difference inside and outside the Dewar results in a different boiling point, which causes the transient vaporization when $LN_2$ ejects out. The all-white color indicates a strong scattering. To get rid of the gas phase, a phase separator is used between the Dewar and the flask.

As a target, the surface smoothness and stability of the flow is key to getting a good signal-to-noise ratio. In fluid mechanics, a flowing liquid can be characterized by Reynolds numbers (*Re*),[19] which is a dimensionless quantity defined as

$$Re = \frac{2\rho v r}{\eta} \qquad (1)$$

where $\rho$, $v$, $r$ and $\eta$ are the density, flow rate, radius and dynamic viscosity of the liquid, respectively. When *Re* < 2300, laminar flow is formed with no lateral mixing or turbulence,[20] in which the liquid is regarded as several layers moving smoothly. Laminar flow offers a good-quality target for optical excitation. From **Eq. (1)**, we know that a low flow rate and a high viscosity are in favor of a small *Re* number. However, as a target designed for the laser with 1 kHz repetition rate, the flow rate (*v*) needs to be greater than 1 m/s to provide each pulse a fresh area. $LN_2$ has a relatively low viscosity ($\eta$ = 150 µPa·s at 77 K) leading to the difficulty in creating a Laminar flow. In our case, to avoid a turbulent flow, *Re* is kept at 2150 by selecting the syringe with appropriate diameter.

It should be noted that the $LN_2$ line is flowing in the ambient environment. This is possible because of the Leidenfrost effect,[21,22] in which an insulating layer is created at the surface by the vaporized $LN_2$ to keep the liquid from boiling rapidly. The liquid line will break into droplets eventually due to the surface energy minimization. The break-up distance $L$ from the needle tip can be calculated by[23]

$$L = 12v\left(\sqrt{\frac{8\rho r^3}{\sigma}} + \frac{6\eta r}{\sigma}\right) \qquad (2)$$

wherein $\sigma$ is the surface tension. For $LN_2$ at 77 K, $\sigma$ is 8.94 mN/m. $L$ = 29 mm is calculated in our case. Before breaking into droplets, the surface smoothness and stability decreases with the distance from the needle tip.

In the optical excitation setup, the laser pulses with an 800 nm central wavelength (370 fs, 0.4 mJ) are focused into the $LN_2$ target. With a lens of 2-inch focal length, the beam waist is about 2.6 μm. The focus position is about 2 mm away from the tip of the syringe needle, which is in the area of a cylindrical line with a smooth surface and far away from droplets. THz waveforms are measured in the forward direction by EOS with a 2-mm thick ZnTe crystal. More details of the experimental setup can be found in Ref. 12.

The normalized THz waveforms from a $LN_2$ line and a water line under the same experimental conditions are shown in **Fig. 2**. Currently, the peak field from $LN_2$ is 0.4 times weaker than that from water. However, the THz signal from $LN_2$ shows a shorter pulse duration. The corresponding spectra without normalization are shown in the inset for comparing the real magnitude between two signals. Under the same excitation and detection conditions, the $LN_2$ shows a broader bandwidth with more high-frequency components. There are two possible reasons. $LN_2$ has a low absorption in THz frequency because it's a nonpolar liquid. Additionally, the vaporized $N_2$ keeps purging the system to preserve the high frequency components. The cutoff frequency is about 2.5 THz, which is limited by the detection crystal. THz wave emission from bulk $LN_2$ under a two-color or double pump excitation was recently reported,[24] in which a bolometer is used for detection. Distinctively, we are using a flowing LN$_2$ line and measuring a temporal THz waveform.

The THz wave emission from liquids is extremely sensitive to the position of the liquid target across the focus. **Fig. 3** shows the cross section of a liquid line and the laser beam. By scanning the position of the LN$_2$ line in *x* direction from *x* = 0, the incident angle ($\alpha$) at the air/liquid interface is continuously increasing from 0° to 90°, which can be calculated by $\alpha = arcsin(\Delta x/r)$. $\Delta x$ is the distance of liquid centroid away from the *z* direction. The laser beam refracts at the air/liquid interface, leading to a deviation from *z* direction. As it has been discussed in our previous work,[25] there is an optimized incident angle to get the maximal THz field coupled out from the liquid, which is determined by the dipole projection in the direction of detection and the total internal reflection of the THz wave at the liquid/air interface.

**Fig. 4a** plots the dependence of THz peak field on the *x*-position of liquid line. By moving the target from negative to positive direction, the electric field has an opposite sign. The maximized signal is obtained at $\Delta x = \pm 170$ μm, showing that the optimal incident angle is 50.6°. As a nonpolar liquid, LN$_2$ has a much lower absorption coefficient (0.8 cm$^{-1}$ at 1 THz)[26] than that of water (220 cm$^{-1}$ at 1 THz)[27]. A lower refractive index is expected, which results in a smaller optimal incident angle. For the laser beam, the difference of refractive index between liquid water (1.3) and LN$_2$ (1.2) is too small to be a dominant factor here. The baseline offset in **Fig. 4a** is the amplitude of THz emission from air plasma when the liquid target is moved away from the laser focus. Then, the field amplitude gradually increases when the liquid line is moved towards the focus. It can be explained the molecular density near the liquid line gradually increased by the vaporization of LN$_2$. Additionally, the result clearly shows that under the excitation of sub-picosecond pulse the signal from a liquid is stronger than that from air.

**Fig. 4b** shows the waveforms when $\Delta x = \pm 170$ μm, respectively. The waveforms have the same shape with an opposite polarity. This is because that the projection of dipole created by electrons has an opposite direction, which clearly indicates that the THz signal is from a liquid phase rather than a vaporized gas phase. This demonstration also shows that the flowing liquid target can be applied to both normal and cryogenic liquid. Moreover, flowing liquid nitrogen is a debris-free target, the mist is vaporized immediately without contamination and scattering.

Currently, THz wave generation from liquids under single-color excitation is explained by the ponderomotive force induced dipole. However, many properties remain unclear. Namely, which properties of liquids can lead to a high field and a broad bandwidth? Besides liquid water, two nonpolar liquids have been tested, α-pinene[16] and liquid nitrogen (in this paper). Both of them provide a broader bandwidth of THz signal than that from water, which suggests the significance of the low absorption. On the other hand, a much brighter white light from liquids than that from air plasma is observed in the experiment, which indicates that the liquid does provide more electrons in the ionization process. But the way of making full use of the ionized electrons to improve the generation efficiency needs more theoretical and experimental studies.

In summary, we report a temporal THz waveform from a flowing cryogenic line under the single-color excitation. Comparing to the THz signal from liquid water, a comparable electric field but with a broader bandwidth is observed. Our results show that $LN_2$ has the potential to be a THz source for generating broadband THz pulses without producing debris. Furthermore, ionized $LN_2$ can also emit X-ray. Along with THz ray, they are able to reflect different dynamics of ionized electrons, respectively. Detecting two synchronized pulses (X-ray and THz ray) as well as the white light in the same ionization process is meaningful to portray a full picture of electron dynamics from excitation to recombination. Also, developing flowing liquid target (line/droplets) for high repetition rate lasers will also facilitate studies of laser-liquid interaction in diverse materials.

**Acknowledgments**

The research at the University of Rochester is sponsored by the Army Research Office under Grant No. W911NF-17-1-0428, Air Force Office of Scientific Research under Grant No. FA9550-18-1-0357, and National Science Foundation under Grant No. ECCS-1916068.

**Data availability**

The data that support the findings of this study are available from the corresponding author upon reasonable request.

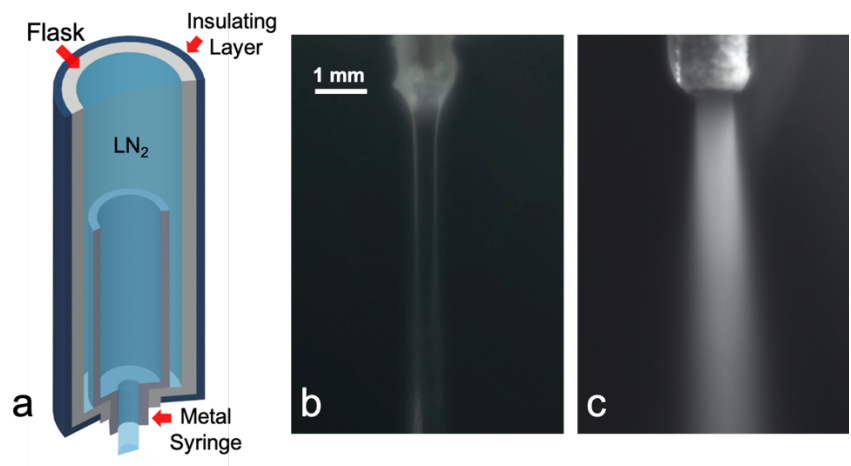

**Fig. 1 a** Diagram of apparatus for guiding a LN$_2$ line. **b** Photo of a flowing LN$_2$ line in ambient environment. **c** Photo of a mixture of gas and liquid nitrogen.

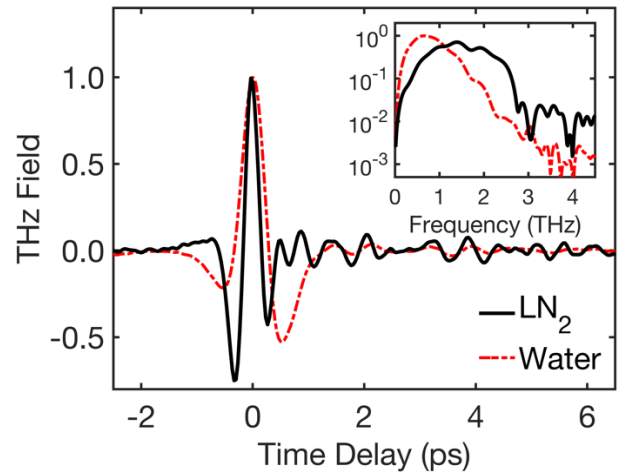

**Fig. 2** Detected THz waveforms from a water line (210 μm) and a LN$_2$ line (400 μm), respectively. The corresponding spectra are shown in the inset. The LN$_2$ signal shows a narrower pulse duration, and its spectrum has a broader bandwidth.

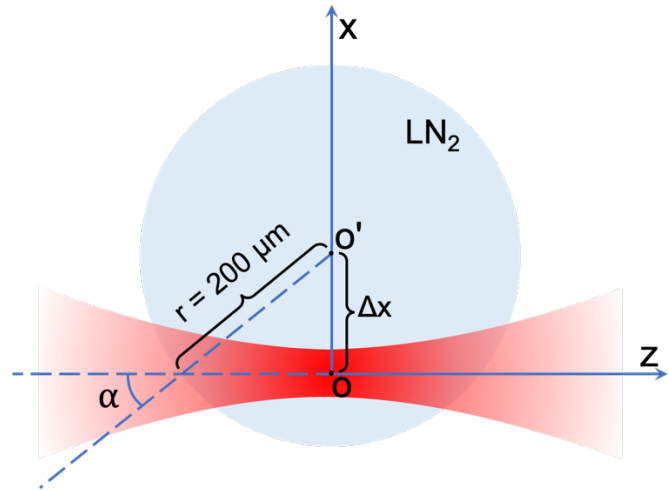

**Fig. 3** The cross section of the liquid line to show the relative position of the LN$_2$ line and the focused laser beam. Laser propagates in $z$ direction. The incident angle $\alpha$ continuously changes by scanning the LN$_2$ line in $x$ direction, in which $\alpha = arcsin(\Delta x/r)$. $\Delta x$ is the distance of liquid centroid away from the $z$ direction. $r$ is the radius of the liquid line.

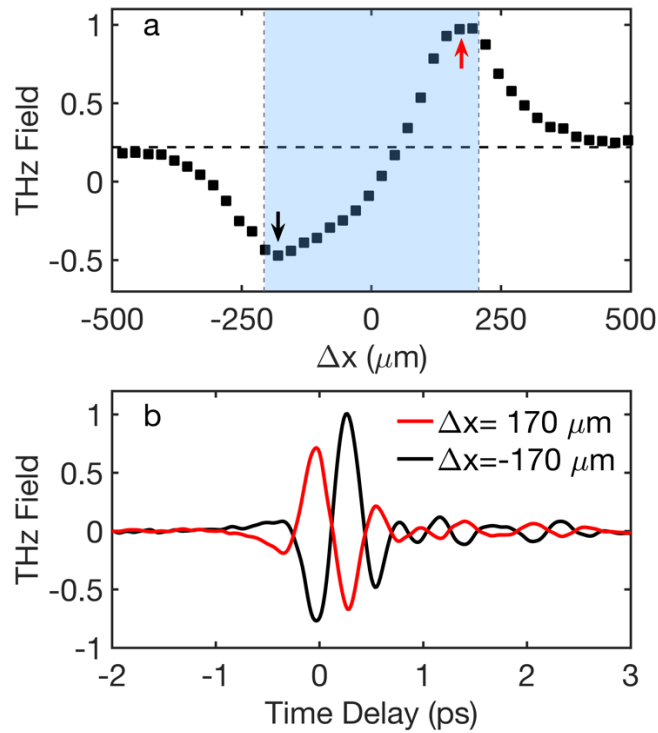

**Fig. 4 a** Dependence of THz field strength on the liquid line position when the time delay is 0 ps. The blue shadow shows the diameter of the liquid line. The baseline above zero shows the amplitude of THz signal from air plasma. **b** Flipped THz waveforms when $\Delta x = -170$ μm and 170 μm, respectively. The corresponding positions are labeled by arrows in plot **a**. The incident angle $\alpha = 50.6°$ when $\Delta x = \pm 170$ μm.

# References

1  Hassan A. Hafez, Sergey Kovalev, Klaas-Jan Tielrooij, Mischa Bonn, Michael Gensch, and Dmitry Turchinovich,  Advanced Optical Materials **8** (3), 1900771 (2020).

2  X.-C. Zhang, Alexander Shkurinov, and Yan Zhang,  Nature Photonics **11** (1), 16 (2017).

3  Tobias Kampfrath, Koichiro Tanaka, and Keith A. Nelson,  Nature Photonics **7** (9), 680 (2013).

4  Xian Li, Tian Qiu, Jiahao Zhang, Edoardo Baldini, Jian Lu, Andrew M. Rappe, and Keith A. Nelson,  Science **364** (6445), 1079 (2019).

5  Hassan A. Hafez, Sergey Kovalev, Jan-Christoph Deinert, Zoltán Mics, Bertram Green, Nilesh Awari, Min Chen, Semyon Germanskiy, Ulf Lehnert, Jochen Teichert et al.,  Nature **561** (7724), 507 (2018).

6  Sharly Fleischer, Yan Zhou, Robert W. Field, and Keith A. Nelson,  Physical Review Letters **107** (16), 163603 (2011).

7  H. Hirori, A. Doi, F. Blanchard, and K. Tanaka,  Applied Physics Letters **98** (9), 091106 (2011).

8  Christoph P. Hauri, Clemens Ruchert, Carlo Vicario, and Fernando Ardana,  Applied Physics Letters **99** (16), 161116 (2011).

9  Guoqian Liao, Yutong Li, Hao Liu, Graeme G Scott, David Neely, Yihang Zhang, Baojun Zhu, Zhe Zhang, Chris Armstrong, and Egle Zemaityte,  Proceedings of the National Academy of Sciences **116** (10), 3994 (2019).

10  L. Malmqvist, L. Rymell, M. Berglund, and H. M. Hertz,  Review of Scientific Instruments **67** (12), 4150 (1996).

11    K. M. George, J. T. Morrison, S. Feister, G. K. Ngirmang, J. R. Smith, A. J. Klim, J. Snyder, D. Austin, W. Erbsen, K. D. Frische et al., High Power Laser Science and Engineering **7**, e50 (2019).

12    Qi Jin, Yiwen E, Kaia Williams, Jianming Dai, and X.-C. Zhang, Applied Physics Letters **111** (7), 071103 (2017).

13    Indranuj Dey, Kamalesh Jana, Vladimir Yu Fedorov, Anastasios D. Koulouklidis, Angana Mondal, Moniruzzaman Shaikh, Deep Sarkar, Amit D. Lad, Stelios Tzortzakis, Arnaud Couairon et al., Nature Communications **8** (1), 1184 (2017).

14    Liang-Liang Zhang, Wei-Min Wang, Tong Wu, Shi-Jia Feng, Kai Kang, Cun-Lin Zhang, Yan Zhang, Yu-Tong Li, Zheng-Ming Sheng, and Xi-Cheng Zhang, Physical Review Applied **12** (1), 014005 (2019).

15    Yuqi Cao, Yiwen E, Pingjie Huang, and X-C Zhang, Applied Physics Letters **117** (2020).

16    Yiwen E, Qi Jin, and X.-C. Zhang, Applied Physics Letters **115** (2019).

17    Evgenia A Ponomareva, Anton N Tcypkin, Semen V Smirnov, Sergey E Putilin, E Yiwen, Sergei A Kozlov, and Xi-Cheng Zhang, Opt. Express **27** (22), 32855 (2019).

18    Qi Jin, Yiwen E, Shenghan Gao, and Xi-Cheng Zhang, Advanced Photonics **2** (1), 015001 (2020).

19    Osborne Reynolds, Philosophical Transactions of the Royal Society of London **174**, 935 (1883).

20    Allan D Kraus, James R Welty, and Abdul Aziz, *Introduction to thermal and fluid engineering*. (CRC Press, 2011).

21    Johann Gottlob Leidenfrost, *De aquae communis nonnullis qualitatibus tractatus*. (Ovenius, 1756).

22      B. S. Gottfried, C. J. Lee, and K. J. Bell,  International Journal of Heat and Mass Transfer **9** (11), 1167 (1966).

23      M. J. McCarthy and N. A. Molloy,  The Chemical Engineering Journal **7** (1), 1 (1974).

24      Alexei V. Balakin, Jean-Louis Coutaz, Vladimir A. Makarov, Igor A. Kotelnikov, Yan Peng, Peter M. Solyankin, Yiming Zhu, and Alexander P. Shkurinov,  Photon. Res. **7** (6), 678 (2019).

25      Yiwen E, Qi Jin, Anton Tcypkin, and XC Zhang,  Appl. Phys. Lett. **113**, 181103 (2018).

26      Jannis Samios, Uwe Mittag, and Thomas Dorfmüller,  Molecular Physics **56** (3), 541 (1985).

27      Tianwu Wang, Pernille Klarskov, and Peter Uhd Jepsen,  IEEE Transactions on Terahertz Science and Technology **4** (4), 425 (2014).